\documentclass[aps,prl,twocolumn]{revtex4} 

\begin{document}

{\bf \noindent Comment on ``Do Intradot Electron-Electron Interactions
Induce Dephasing?''}
\\
\\
In a recent Letter, Jiang, Sun, Xie and Wang \cite{jiang} study transport
through an interacting quantum dot embedded in one arm of an 
Aharonov-Bohm interferometer.
Based on a theoretical analysis of the Aharonov-Bohm oscillation amplitude, 
Jiang {\it et al.} claim, contrary to earlier work by two of us \cite{kg},
that at finite temperature the intradot interaction will {\em not} lead to any
dephasing. 
Likewise, they claim that the theoretically predicted \cite{kg} and 
experimentally verified \cite{kobayashi} asymmetry of the 
Aharonov-Bohm oscillation amplitude is {\em not} associated with dephasing.
In this Comment, we point out severe inconsistencies in the analysis of 
Ref.~\onlinecite{jiang}, and show that their conclusions are ill-founded.

Our main point is that the authors of Ref.~\onlinecite{jiang} employ an
approximation scheme for the Green's functions, which, {\em by construction}, 
is inappropriate to describe spin-flip-related dephasing \cite{kg} or
any other inelastic process that is caused by intradot interaction.
To evaluate the Green's function of the quantum dot they consider the 
uncoupled dot, and then assign by hand constant widths to the 
bare many-body dot levels, as described in the paragraph below Eq.~(3) 
of Ref.~\cite{jiang}.
This procedure results in the symmetrized Hartree approximation
(a low-hierarchy variant of the equation-of-motion scheme).
The latter is an effective {\it single-particle approximation} that 
oversimplifies the role of interaction.
The failure of this scheme is most strikingly demonstrated in the first
paragraph of the right column on page 3 of Ref.~\onlinecite{jiang}, where 
the authors' ``proof'' of the nonexistence of dephasing relies on the
relation $T_d = \Gamma_{11}^s \Gamma_{44}^d |\tilde G_{41}^r|^2$ for
the transmission probability through the arm containing the quantum dot.
This is equivalent to the relation $T=|t|^2$, valid for
{\em non-interacting} systems, where $t$ is the transmission amplitude
(proportional to the single-particle Green's function involving one source- 
and one drain-electron operator).
For {\em interacting} systems with inelastic channels, 
however, this relation is no longer valid,
as has been shown in the literature \cite{mw,kg}.

To illustrate the importance of this point for the present context, 
we explicitly write down the cotunneling transmission of
an electron at the Fermi energy through a single-level quantum dot with level 
energy $\epsilon$, tuned away from resonance, and charging energy 
$U\rightarrow \infty$.
The correct employment of the many-body formalism \cite{mw,kg} yields
\begin{equation}
\label{general}
  T_{\rm dot} = -\frac{2\Gamma_{\rm L}\Gamma_{\rm R}}
  {\Gamma} {\rm Im} \, G_{\rm dot}^{\rm ret} \, ,
\end{equation}
as opposed to the result
\begin{equation}
\label{non}
  T_{\rm dot}^{U=0} = \Gamma_{\rm L}\Gamma_{\rm R}
  |G_{\rm dot}^{\rm ret}|^2 \, , 
\end{equation}
that is valid for {\em non-interacting} electrons only.
Here, $\Gamma_{\rm L}$ and $\Gamma_{\rm R}$ measure the coupling strengths 
between dot and the left and right lead, respectively, and 
$\Gamma=\Gamma_{\rm L}+\Gamma_{\rm R}$ is the broadening of the dot level.
The {\em correct} cotunneling transmission \cite{kg} is thus 
obtained from Eq.~(\ref{general}) as
\begin{equation}
\label{correct}
  T_{\rm dot}^{\rm correct} = 
  \frac{\Gamma_{\rm L} \Gamma_{\rm R}}{\epsilon^2} \, ,
\end{equation}
while within the Hartree approximation, Eq.~(\ref{general}) and (\ref{non}) 
are equivalent and lead to the {\em wrong} result
\begin{equation}
\label{wrong}
  T_{\rm dot}^{\rm Hartree} = \frac{1}{[1+f(\epsilon)]^2} \cdot 
  \frac{\Gamma_{\rm L} \Gamma_{\rm R}} {\epsilon^2} \, ,
\end{equation}
in order $\Gamma^2$.
Equations~(\ref{correct}) and (\ref{wrong}) differ by a factor 
$1/[1+f(\epsilon)]^2$, where $f(\epsilon)$ is the Fermi function.
Repeating the above-mentioned discussion of the ``decoherence rate'' $r_T$
on page 3 of Ref.~\onlinecite{jiang} but with the correct expression for 
$T_{\rm dot}$ immediately leads to $r_T \rightarrow 1$ for $\epsilon > 0$ 
but $r_T \rightarrow 1/2$ for $\epsilon < 0$, indicating interaction-induced
dephasing (and consequently asymmetry) for both a two-terminal and an open
geometry.  
This is in accordance with Ref.~\onlinecite{kg}. 
Qualitatively similar results hold at resonance.

There are several other inconsistencies in the paper.
We give here some examples: 
(1) While $r_T$ and $r_G$ are claimed to be close to each other at low 
$k_B T$, $r_G$ clearly exceeds 1 near resonance (cf. Fig. 2 of 
Ref.~\onlinecite{jiang}), invalidating it as a good measure of coherence or 
"visibility".
(2) Equation~(4) of Ref.~\onlinecite{jiang} is wrong, as it relies on the
single-particle formalism.
(3) The quantity $\Delta G(\phi)$ should, by construction, vanish 
for $\phi=0$, which is clearly not the case in Fig. 2b of 
Ref.~\onlinecite{jiang}.

This work was supported by the Deutsche Forschungsgemeinschaft through
SFB 491 and GRK 726, the US-Israel BSF, the ISF of the Israel 
Academy of Science, the Alexander von Humboldt foundation, the 
EC HPRN-CT-2002-00302-RTN, and the NSF grant DMR 0210575.

\vspace*{0.2cm} 
 
\noindent J\"urgen K\"onig$^1$, Yuval Gefen$^2$, and Alessandro Silva$^{3}$
 
{\small $^1$Institut f\"ur Theoretische Physik III\\
\hspace*{.5cm}Ruhr-Universit\"at Bochum\\
\hspace*{.5cm}44780 Bochum, Germany

$^2$Department of Condensed Matter Physics\\
\hspace*{.5cm}The Weizmann Institute of Science\\
\hspace*{.5cm}76100 Rehovot, Israel

$^3$Center for Materials Theory\\
\hspace*{.5cm}Department of Physics and Astronomy\\
\hspace*{.5cm}Rutgers University, Piscataway, NJ 08854, USA}

\vspace*{0.5cm} 
 
 
 

\end{document}